\documentstyle[11pt,aaspp4,epsf]{article}
\def\bc {\begin{center}}
\def\ec {\end{center}}

\def\bn {\begin{enumerate}}
\def\en {\end{enumerate}}

\def\be {\begin{equation}}
\def\ee {\end{equation}}

\def\bea{\begin{eqnarray}}
\def\eea{\end{eqnarray}}

\def\vs{{\it vs.}}
\def\etal{{\it et al.}}

\newcommand{\NEP}{\mbox{NEP}}
\newcommand{\mW}{\mbox{{\bf W}}}
\newcommand{\mP}{\mbox{{\bf P}}}
\newcommand{\mT}{\mbox{{\bf T}}}
\newcommand{\mA}{\mbox{{\bf A}}}
\newcommand{\mS}{\mbox{{\bf S}}}
\newcommand{\mB}{\mbox{{\bf B}}}
\newcommand{\mone}{\mbox{{\bf 1}}}
\newcommand{\mI}{\mbox{{\bf I}}}
\newcommand{\mX}{\mbox{{\bf X}}}
\newcommand{\mY}{\mbox{{\bf Y}}}
\begin{document}

\slugcomment{Presented 22 Nov 96 at the IAS CMB Data Analysis Workshop}

\title{Scanning and Mapping Strategies for CMB Experiments}
\author{E. L. Wright}
\authoraddr{UCLA Physics \& Astronomy}

\begin{abstract}
CMB anisotropy experiments seeking to make maps with more pixels
than the 6144 pixels used by the {\sl COBE} DMR need to address
the practical issues of the computer time and storage required
to make maps.  A simple, repetitive scan pattern reduces these
requirements but leaves the experiment vulnerable to systematic
errors and striping in the maps.  In this paper I give a time-ordered
method for map-making with one-horned experiments that has reasonable
memory and CPU needs but can handle complex {\sl COBE}-like scans
paths and $1/f$ noise.
\end{abstract}

\section{Introduction}

Many new CMB missions such as {\sl COBRAS/SAMBA} have adopted scan 
patterns with a relatively small number of scan circles in order to
simplify the data analysis.  The very successful {\sl COBE} pattern
of cycloidal scans that cross through each pixel in many different
directions is not used, and this leads to an increased sensitivity
to systematic errors.  These new missions have adopted a total power
or one-horned approach to maximize sensitivity relative to the 
differential approach used by {\sl COBE}.  But this often leads to
excess low frequency or $1/f$ noise.

\section{Direct Imaging}

A simple scan pattern with great circle scans perpendicular to the
Sun line can lead to {\em striping} even in the absence of $1/f$ noise.
The simplest reference pattern
is to reference each pixel in the scan to the NEP.  Thus the
equation to be solved in the map-making for pixel $j$ which is
the $i^{th}$ pixel in the $k^{th}$ scan is
\be
T_j - T_{\NEP} = S(i,k) - S(\NEP,k).
\ee
We assume that the samples $S(i,k)$ are uncorrelated (white noise)
except for a constant baseline that has to be determined for each
scan.  But the baseline cancels out in this equation.
The noise associated with this equation is $\sqrt{2}$ times the
noise per pixel because two values are subtracted.  

These observations can be inverted directly giving
\bea
T_{\NEP} & = & 0 \nonumber\\
T_j & = & S(i,k) - S(\NEP,k) \nonumber\\
T_{\mbox{SEP}} & = & \frac{2}{N} 
\sum_k \left[S(\mbox{SEP},k) - S(\NEP,k)\right] \\
\eea
where the first line is an assumption for $T_{\NEP}$.
This appears to be the approach described in the text of 
Janssen \etal\markcite{JSWSL96} (1996).

The covariance matrix can also be computed directly:
\bea
\left\langle T_{\mbox{SEP}} T_{\mbox{SEP}} \right\rangle & = &
\frac{4 \sigma_1^2}{N} \nonumber\\
\left\langle T_i T_{\mbox{SEP}} \right\rangle & = &
\frac{2\sigma_1^2}{N} \nonumber\\
\left\langle T_i T_i \right\rangle & = & 2 \sigma_1^2 \nonumber\\
\left\langle T_i T_j \right\rangle & = &
\cases{\sigma_1^2,&if on the same scan;\cr 0,&otherwise.\cr}
\\
\eea
The non-zero covariance of pixels on the same scan circle shows
the presence of striping.  The factor of 2 in the variance of
$T_i$ shows that this method has lost the putative $\sqrt{2}$
advantage of one-horned systems.

How can the $\sqrt{2}$ be recovered?  A better baseline estimator
is needed.  For this white noise case, the average of all the
noises in a scan circle is the optimum baseline estimator.
But we don't have the noises -- we only measure samples which are
the sum of noise plus sky.  Therefore we need to do any iterative
solution to find the baseline:  
\bn
\item subtract the signal from the scan using an estimate of the map,
\item compute the optimum baseline estimate, and
\item use the signal minus baseline to construct a new estimate of
the map.
\en
This procedure is almost identical to the time-ordered iterative
approach used in Wright, Hinshaw \& Bennett \markcite{WHB96} (1996).  
The only
difference is that for differential data the contribution of the
map to the ``baseline'' is just minus the sky temperature in the
reference beam.

If applied to a model with $N/2$ scans of $N$ pixels crossing only
at the NEP and SEP, the optimum approach reduces the sensitivity
loss to a factor of $\sqrt{3/2}$ compared to an ideal total power
system in the large $N$ limit.  For large $N$, the difference between
the NEP and SEP is determined to great precision, so the only
significant terms in the covariance matrix are
\bea
\left\langle T_i T_i \right\rangle & = & \frac{3}{2} \sigma_1^2 \nonumber\\
\left\langle T_i T_j \right\rangle & = &
\cases{\frac{1}{2}\sigma_1^2,&if on the same scan;\cr 0,&otherwise.\cr}
\\
\eea
This still has stripes, but both the noise and striping are reduced.
In a real sky map, the number of multiply observed pixels in the
polar caps is quite large, so the both the excess noise and the striping
are reduced even more.  This approach of subtracting an optimal baseline
after correcting the baseline iteratively for the effect of the map on the
baseline is equivalent to the method of adjusting the baselines in each scan
circle using the set of overlapping pixels that was used by {\sl FIRS}
(Meyer, Cheng \& Page \markcite{MCP91} 1991) and 
is planned by {\sl COBRAS/SAMBA}
(Bersanelli \etal\markcite{BBEGL96} 1996).

\begin{figure}[tbp]
\plotone{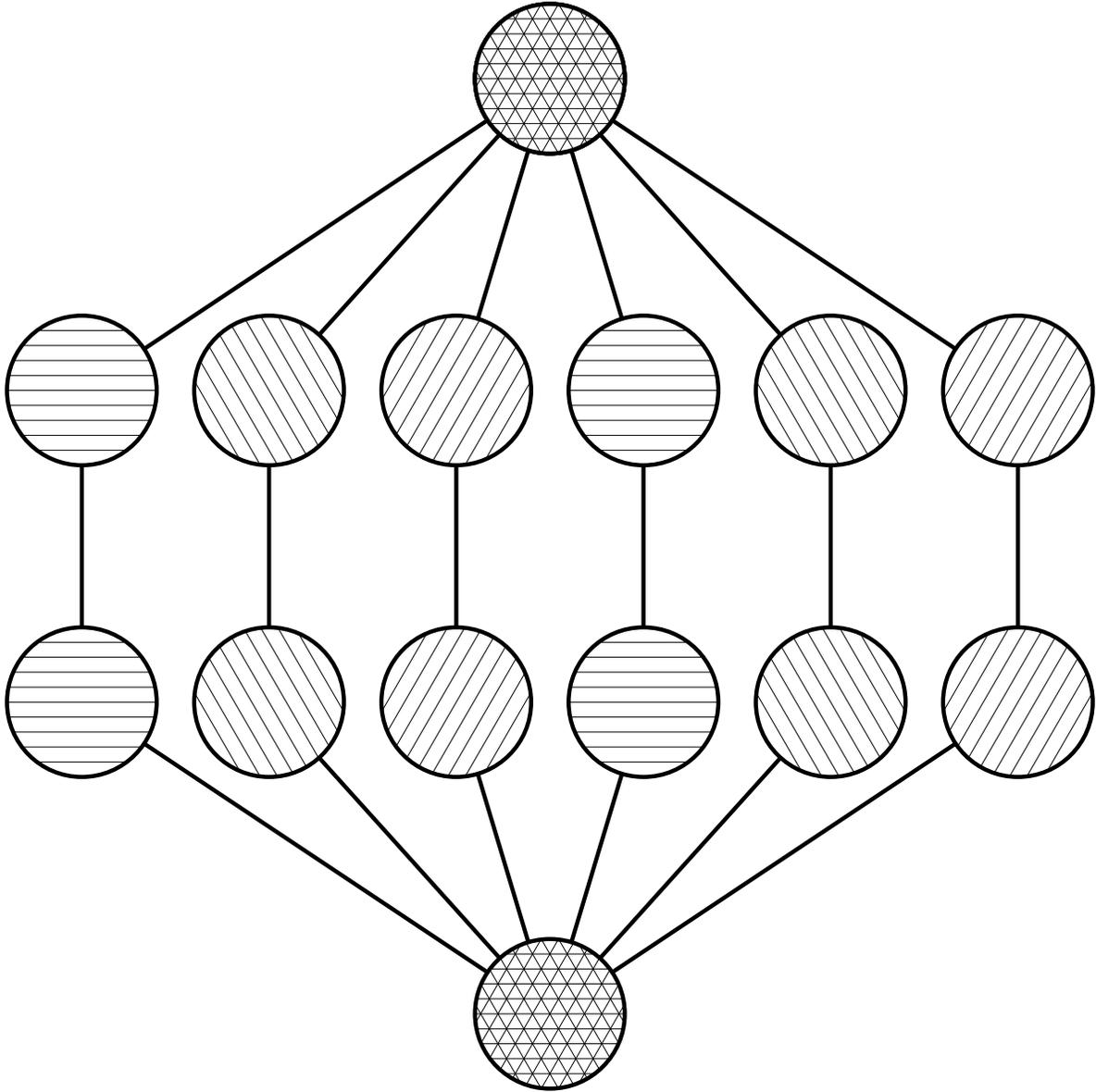}
\caption{The toy sky model with 14 pixels observed in 3 scan circles,
shown with different cross-hatching, totaling 18 data points.
\label{fig:toy_sky.eps}}
\end{figure}

\section{Toy Model}

To illustrate the operations in the map-making process, I will
use a toy model of the sky.  This toy model has
$N/2$ scans each with $N$ pixels crossing only at the NEP (North
Ecliptic Pole) and SEP.  Figure \ref{fig:toy_sky.eps}
shows the toy model with $N = 6$, so there are $n_d = N^2/2 = 18$ data 
points and $n_p = (N/2)(N-2)+2 = 14$ pixels.

The pixel observed during each data sample must be known.
Let $p(j)$ be the pixel observed during the $j^{th}$ data
sample.
I can write this as a matrix by defining \mP\ to be an
$n_d \times n_p$ matrix, with 1 row for each observation and 1 column for
each pixel.
\mP\ is all zero except for a single ``1'' in each row at the column
observed for that row's datum.
Note that \mP\ is the operation of ``flying the mission through a skymap''
and that \mP\ maps a skymap into a data stream.
The matrix \mP\ for the toy model with $N = 6$ is given by
\be
\mP = \left(
\begin{array}{rrrrrrrrrrrrrr}
   1 &   0 &   0 &   0 &   0 &   0 &   0 &   0 &   0 &   0 &   0 &   0 &   0 &   0 \\
   0 &   1 &   0 &   0 &   0 &   0 &   0 &   0 &   0 &   0 &   0 &   0 &   0 &   0 \\
   0 &   0 &   1 &   0 &   0 &   0 &   0 &   0 &   0 &   0 &   0 &   0 &   0 &   0 \\
   0 &   0 &   0 &   0 &   0 &   0 &   0 &   0 &   0 &   0 &   0 &   0 &   0 &   1 \\
   0 &   0 &   0 &   1 &   0 &   0 &   0 &   0 &   0 &   0 &   0 &   0 &   0 &   0 \\
   0 &   0 &   0 &   0 &   1 &   0 &   0 &   0 &   0 &   0 &   0 &   0 &   0 &   0 \\
   1 &   0 &   0 &   0 &   0 &   0 &   0 &   0 &   0 &   0 &   0 &   0 &   0 &   0 \\
   0 &   0 &   0 &   0 &   0 &   1 &   0 &   0 &   0 &   0 &   0 &   0 &   0 &   0 \\
   0 &   0 &   0 &   0 &   0 &   0 &   1 &   0 &   0 &   0 &   0 &   0 &   0 &   0 \\
   0 &   0 &   0 &   0 &   0 &   0 &   0 &   0 &   0 &   0 &   0 &   0 &   0 &   1 \\
   0 &   0 &   0 &   0 &   0 &   0 &   0 &   1 &   0 &   0 &   0 &   0 &   0 &   0 \\
   0 &   0 &   0 &   0 &   0 &   0 &   0 &   0 &   1 &   0 &   0 &   0 &   0 &   0 \\
   1 &   0 &   0 &   0 &   0 &   0 &   0 &   0 &   0 &   0 &   0 &   0 &   0 &   0 \\
   0 &   0 &   0 &   0 &   0 &   0 &   0 &   0 &   0 &   1 &   0 &   0 &   0 &   0 \\
   0 &   0 &   0 &   0 &   0 &   0 &   0 &   0 &   0 &   0 &   1 &   0 &   0 &   0 \\
   0 &   0 &   0 &   0 &   0 &   0 &   0 &   0 &   0 &   0 &   0 &   0 &   0 &   1 \\
   0 &   0 &   0 &   0 &   0 &   0 &   0 &   0 &   0 &   0 &   0 &   1 &   0 &   0 \\
   0 &   0 &   0 &   0 &   0 &   0 &   0 &   0 &   0 &   0 &   0 &   0 &   1 &   0 
\end{array}\right)
\ee
The NEP is pixel 1, the SEP is pixel 14, and scan circle 1 is 
pixels 1, 2, 3, 14, 4, 5 in order, while scan circle 2 is pixels
1, 6, 7, 14, 8, 9.

Define the
matrix \mW\ to be an $n_d \times n_d$ matrix that corrects the data
for the optimal baseline.  For the white noise case of the toy model,
this matrix is given by
\be
\mW = \frac{1}{5}\left(
\begin{array}{rrrrrrrrrrrrrrrrrr}
   5 &  -1 &  -1 &  -1 &  -1 &  -1 &   0 &   0 &   0 &   0 &   0 &   0 &   0 &   0 &   0 &   0 &   0 &   0\\
  -1 &   5 &  -1 &  -1 &  -1 &  -1 &   0 &   0 &   0 &   0 &   0 &   0 &   0 &   0 &   0 &   0 &   0 &   0\\
  -1 &  -1 &   5 &  -1 &  -1 &  -1 &   0 &   0 &   0 &   0 &   0 &   0 &   0 &   0 &   0 &   0 &   0 &   0\\
  -1 &  -1 &  -1 &   5 &  -1 &  -1 &   0 &   0 &   0 &   0 &   0 &   0 &   0 &   0 &   0 &   0 &   0 &   0\\
  -1 &  -1 &  -1 &  -1 &   5 &  -1 &   0 &   0 &   0 &   0 &   0 &   0 &   0 &   0 &   0 &   0 &   0 &   0\\
  -1 &  -1 &  -1 &  -1 &  -1 &   5 &   0 &   0 &   0 &   0 &   0 &   0 &   0 &   0 &   0 &   0 &   0 &   0\\
   0 &   0 &   0 &   0 &   0 &   0 &   5 &  -1 &  -1 &  -1 &  -1 &  -1 &   0 &   0 &   0 &   0 &   0 &   0\\
   0 &   0 &   0 &   0 &   0 &   0 &  -1 &   5 &  -1 &  -1 &  -1 &  -1 &   0 &   0 &   0 &   0 &   0 &   0\\
   0 &   0 &   0 &   0 &   0 &   0 &  -1 &  -1 &   5 &  -1 &  -1 &  -1 &   0 &   0 &   0 &   0 &   0 &   0\\
   0 &   0 &   0 &   0 &   0 &   0 &  -1 &  -1 &  -1 &   5 &  -1 &  -1 &   0 &   0 &   0 &   0 &   0 &   0\\
   0 &   0 &   0 &   0 &   0 &   0 &  -1 &  -1 &  -1 &  -1 &   5 &  -1 &   0 &   0 &   0 &   0 &   0 &   0\\
   0 &   0 &   0 &   0 &   0 &   0 &  -1 &  -1 &  -1 &  -1 &  -1 &   5 &   0 &   0 &   0 &   0 &   0 &   0\\
   0 &   0 &   0 &   0 &   0 &   0 &   0 &   0 &   0 &   0 &   0 &   0 &   5 &  -1 &  -1 &  -1 &  -1 &  -1\\
   0 &   0 &   0 &   0 &   0 &   0 &   0 &   0 &   0 &   0 &   0 &   0 &  -1 &   5 &  -1 &  -1 &  -1 &  -1\\
   0 &   0 &   0 &   0 &   0 &   0 &   0 &   0 &   0 &   0 &   0 &   0 &  -1 &  -1 &   5 &  -1 &  -1 &  -1\\
   0 &   0 &   0 &   0 &   0 &   0 &   0 &   0 &   0 &   0 &   0 &   0 &  -1 &  -1 &  -1 &   5 &  -1 &  -1\\
   0 &   0 &   0 &   0 &   0 &   0 &   0 &   0 &   0 &   0 &   0 &   0 &  -1 &  -1 &  -1 &  -1 &   5 &  -1\\
   0 &   0 &   0 &   0 &   0 &   0 &   0 &   0 &   0 &   0 &   0 &   0 &  -1 &  -1 &  -1 &  -1 &  -1 &   5\\
\end{array}
\right)
\ee

The matrix \mW\ is the $n_d \times n_d$ matrix that applies the
optimal baseline correction.  Then for an observed data stream
\mS, the desired map is the
\mT\ that best satisfies the $n_d$ equations:
\be
\mW\mP\mT = \mW\mS
\ee
The application of \mW\ to \mS\ makes the elements of the right hand side
uncorrelated.  
Actually there is still an ${\cal O}(1/N)$ correlation because the
true mean of the scan is completely undetermined, but I will ignore
this effect which is negligible for real CMB experiments.
Let $\sigma$ be the standard deviation of the noise in these
values, which I will take to be $\sigma = 1$ in the following examples.
Note that this implies that the standard deviation of the original data is 
somewhat less than one due to the variance introduced by baseline subtraction.
Proceeding in the normal least squares fashion, make $n_p$ normal equations:
\be
\mP^T\mW^T\frac{1}{\sigma^2}\mW\mP\mT = \mP^T\mW^T\frac{1}{\sigma^2}\mW\mS
\ee
The matrix $\mP^T$ is the operation of ``summing a data stream into pixels''.
Thus the $n_p \times 1$ right hand side is 
$$\mB = \mP^T\mW^T\frac{1}{\sigma^2}\mW\mS$$
and the $n_p \times n_p$ correlation matrix is
$$\mA = \mP^T\mW^T\frac{1}{\sigma^2}\mW\mP$$
and the equation to solve is
\be
\mA\mT = \mB
\ee
For large problems
this should be solved iteratively, and only products of 
the form \mA\ times a vector are needed.

However, for the toy model $\mA^{-1}$ can be computed directly.
Of course, \mA\ is singular so an extra equation stating that
the sum of the map is zero must be added to the $n_d$ equations
from the observations.  This adds the matrix \mone\
which is matrix with all elements equal to unity to \mA.
Do not confuse \mone\ with the identity matrix \mI.
The generalized inverse of \mA\ is then given by
\be
\mA^{-1} = \left(\mA + \mone\right)^{-1} - n_p^{-2} \mone.
\ee
This particular form for the generalized inverse is a special case that
only fixes the zero eigenvalue associated with the mean of the map.
For experiments with partial sky coverage or disconnected regions
the more general form of the generalized inverse should be used:
\be
\mA^{-1} \approx \lim_{\epsilon \rightarrow 0^+}
\left(\mA + \epsilon\mI\right)^{-1}.
\ee
When solving $\mA\mX = \mY$ be sure that \mY\ is orthogonal to
all the eigenvectors of \mA\ with zero eigenvalues.
For the toy model
\be
\mA = \frac{1}{25}\left(
\begin{array}{rrrrrrrrrrrrrr}
  90 &  -6 &  -6 &  -6 &  -6 &  -6 &  -6 &  -6 &  -6 &  -6 &  -6 &  -6 &  -6 & -18\\
  -6 &  30 &  -6 &  -6 &  -6 &   0 &   0 &   0 &   0 &   0 &   0 &   0 &   0 &  -6\\
  -6 &  -6 &  30 &  -6 &  -6 &   0 &   0 &   0 &   0 &   0 &   0 &   0 &   0 &  -6\\
  -6 &  -6 &  -6 &  30 &  -6 &   0 &   0 &   0 &   0 &   0 &   0 &   0 &   0 &  -6\\
  -6 &  -6 &  -6 &  -6 &  30 &   0 &   0 &   0 &   0 &   0 &   0 &   0 &   0 &  -6\\
  -6 &   0 &   0 &   0 &   0 &  30 &  -6 &  -6 &  -6 &   0 &   0 &   0 &   0 &  -6\\
  -6 &   0 &   0 &   0 &   0 &  -6 &  30 &  -6 &  -6 &   0 &   0 &   0 &   0 &  -6\\
  -6 &   0 &   0 &   0 &   0 &  -6 &  -6 &  30 &  -6 &   0 &   0 &   0 &   0 &  -6\\
  -6 &   0 &   0 &   0 &   0 &  -6 &  -6 &  -6 &  30 &   0 &   0 &   0 &   0 &  -6\\
  -6 &   0 &   0 &   0 &   0 &   0 &   0 &   0 &   0 &  30 &  -6 &  -6 &  -6 &  -6\\
  -6 &   0 &   0 &   0 &   0 &   0 &   0 &   0 &   0 &  -6 &  30 &  -6 &  -6 &  -6\\
  -6 &   0 &   0 &   0 &   0 &   0 &   0 &   0 &   0 &  -6 &  -6 &  30 &  -6 &  -6\\
  -6 &   0 &   0 &   0 &   0 &   0 &   0 &   0 &   0 &  -6 &  -6 &  -6 &  30 &  -6\\
 -18 &  -6 &  -6 &  -6 &  -6 &  -6 &  -6 &  -6 &  -6 &  -6 &  -6 &  -6 &  -6 &  90
\end{array}
\right)
\ee
The covariance matrix is given by $\mA^{-1} = $ 
$$
{\scriptsize
\left(
\begin{array}{rrrrrrrrrrrrrr}
  0.24&-0.02&-0.02&-0.02&-0.02&-0.02&-0.02&-0.02&-0.02&-0.02&-0.02&-0.02&-0.02& 0.01\\
 -0.02& 0.87& 0.18& 0.18& 0.18&-0.17&-0.17&-0.17&-0.17&-0.17&-0.17&-0.17&-0.17&-0.02\\
 -0.02& 0.18& 0.87& 0.18& 0.18&-0.17&-0.17&-0.17&-0.17&-0.17&-0.17&-0.17&-0.17&-0.02\\
 -0.02& 0.18& 0.18& 0.87& 0.18&-0.17&-0.17&-0.17&-0.17&-0.17&-0.17&-0.17&-0.17&-0.02\\
 -0.02& 0.18& 0.18& 0.18& 0.87&-0.17&-0.17&-0.17&-0.17&-0.17&-0.17&-0.17&-0.17&-0.02\\
 -0.02&-0.17&-0.17&-0.17&-0.17& 0.87& 0.18& 0.18& 0.18&-0.17&-0.17&-0.17&-0.17&-0.02\\
 -0.02&-0.17&-0.17&-0.17&-0.17& 0.18& 0.87& 0.18& 0.18&-0.17&-0.17&-0.17&-0.17&-0.02\\
 -0.02&-0.17&-0.17&-0.17&-0.17& 0.18& 0.18& 0.87& 0.18&-0.17&-0.17&-0.17&-0.17&-0.02\\
 -0.02&-0.17&-0.17&-0.17&-0.17& 0.18& 0.18& 0.18& 0.87&-0.17&-0.17&-0.17&-0.17&-0.02\\
 -0.02&-0.17&-0.17&-0.17&-0.17&-0.17&-0.17&-0.17&-0.17& 0.87& 0.18& 0.18& 0.18&-0.02\\
 -0.02&-0.17&-0.17&-0.17&-0.17&-0.17&-0.17&-0.17&-0.17& 0.18& 0.87& 0.18& 0.18&-0.02\\
 -0.02&-0.17&-0.17&-0.17&-0.17&-0.17&-0.17&-0.17&-0.17& 0.18& 0.18& 0.87& 0.18&-0.02\\
 -0.02&-0.17&-0.17&-0.17&-0.17&-0.17&-0.17&-0.17&-0.17& 0.18& 0.18& 0.18& 0.87&-0.02\\
  0.01&-0.02&-0.02&-0.02&-0.02&-0.02&-0.02&-0.02&-0.02&-0.02&-0.02&-0.02&-0.02& 0.24
\end{array}
\right)
}
$$
Note that this particular form of the generalized inverse gives a zero sum
for each row of $\mA^{-1}$ and this introduces the anti-correlation
for points on different scan circles.

\section{$1/f$ Noise}

Most instruments do not produce white noise, but have slow drifts and
other anomalies that produce an excess noise at low frequencies in the
output.  The unknown baseline for each scan circle in the toy model
used earlier is an example of low frequency noise: it corresponds to
a $\delta(f)$ spike in the noise power spectrum at zero frequency.
Real instruments have this zero frequency spike, but they also usually
have a more gradual rise in the noise power spectrum as $f \rightarrow 0$.
Typically an excess noise varying like $1/f$ is seen, giving a noise
power spectrum
\be
P(f) = 2\Delta t \, \sigma_1^2 \left(1 + \frac{f_K}{f}\right)
\ee
where $\Delta t$ is the sampling interval, $\sigma_1$ is the noise
in one sample ignoring the $1/f$ term, and $f_K$ is the $1/f$
``knee'' frequency.

\begin{figure}[tbp]
\plotone{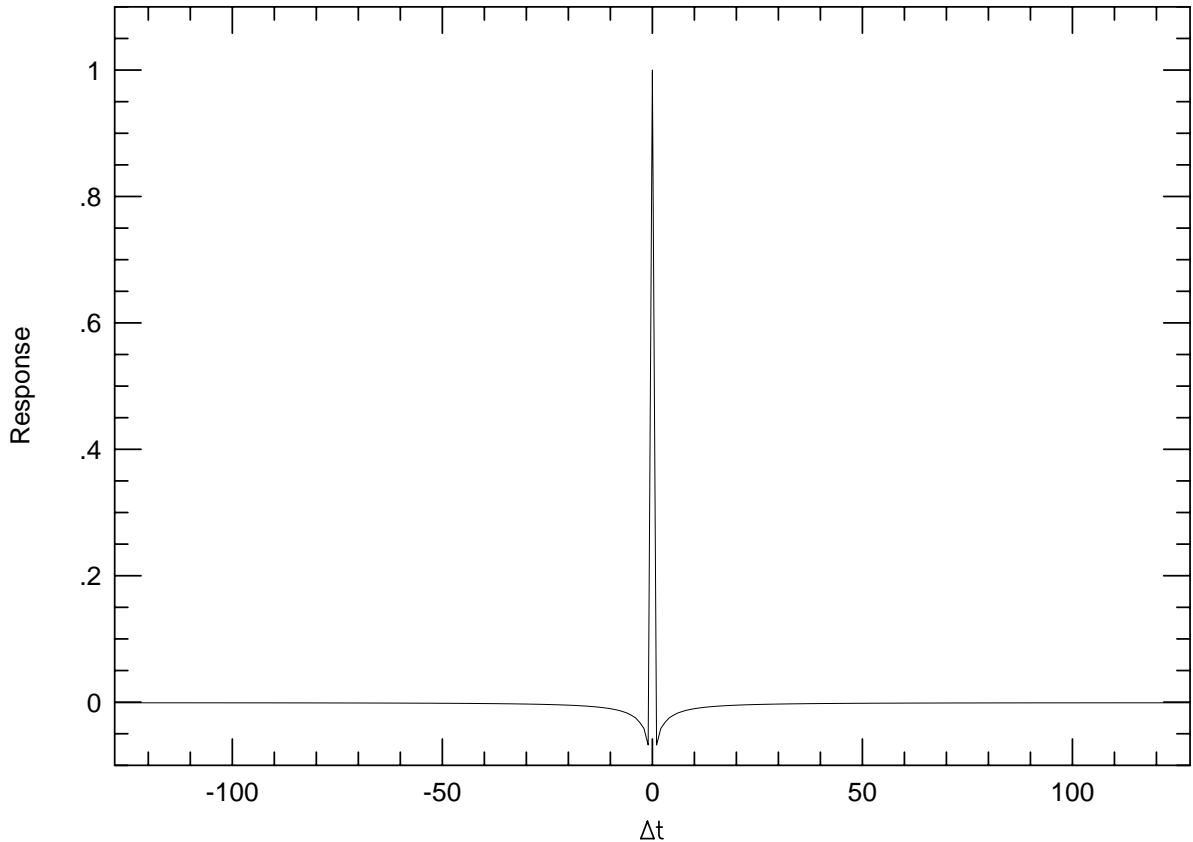}
\caption{The pre-whitening filter for $f_K = 10 f_{spin}$ with
256 points per spin.
\label{fig:w_0p1.eps}}
\end{figure}

In a spin-scanned system, $1/f$ noise is only important if the
spin rate is less than $f_K$.  For {\sl COBRAS/SAMBA}\/ the
spin rate should be faster than the output drifts in either the
bolometers or the differencing HEMTs.  But in balloon-borne
experiments, drifts in the atmosphere will produce excess
low frequency noise that must be correctly treated during
data analysis.

In the presence of $1/f$ noise, successive samples of the 
signal are correlated.  In order to simplify the analysis, a
``pre-whitening'' filter with transmission
\be
W(f) = \sqrt{\frac{f}{f+f_K}}
\ee
should be applied.  After applying this filter, the noise in
different samples will be uncorrelated.  The impulse response
function of $W(f)$ in the time domain will integrate to
zero because the response of $W(f)$ vanishes for $f=0$.
The impulse response function $W(t)$ will have a $\delta(t)$ spike
at zero time because $W(f) \rightarrow 1$ as $f \rightarrow \infty.$
Thus the action of $W$ is to replace the signal stream with
signal minus baseline, and the baseline used is the optimal
baseline estimator.  Figure \ref{fig:w_0p1.eps} show the
pre-whitening filter for $f_K = 10 f_{spin}$.

While the noise in the output of $W$ is uncorrelated,
this filter makes each observation depend on many sky values.
For a discretely sampled data, the $W$ in the time domain
is given by a vector of weights $W_k$, where $W_0 = 1$
and $\sum W_k = 0$.  For {\it post facto} data analysis the
time symmetric filter with $W_{-k} = W_k$ can be used
instead of a causal filter.

\begin{figure}[tbp]
\plotone{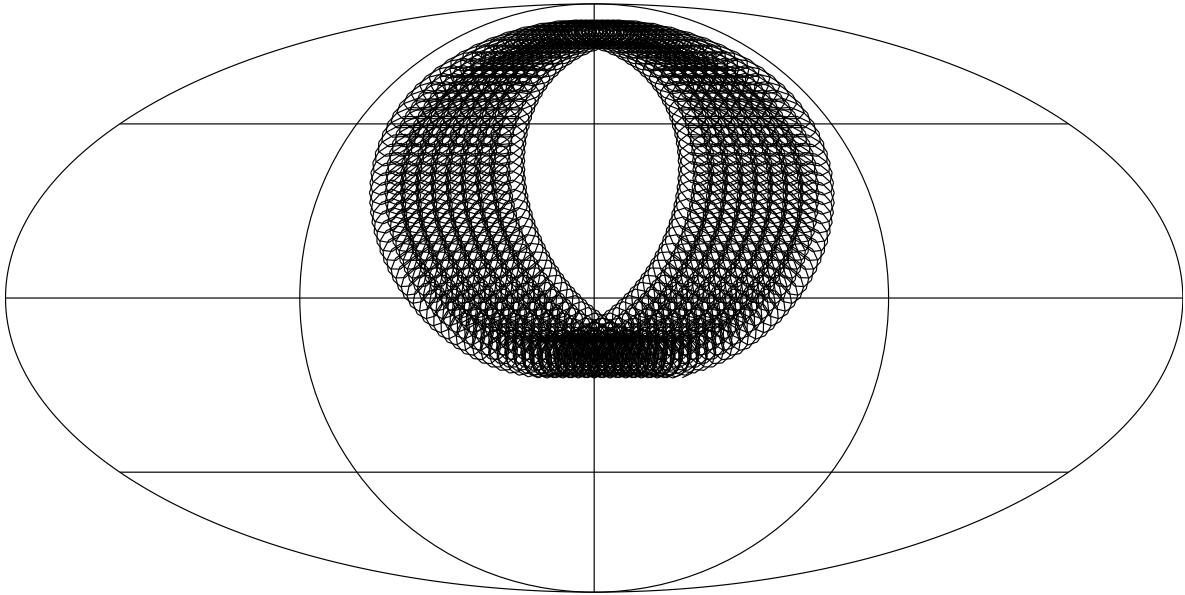}
\caption{A complex scan pattern with a $10^\circ$ circular scan superimposed
on rotation around the zenith and the diurnal rate.
\label{fig:ucsbscan.eps}}
\end{figure}

\begin{figure}[tbp]
\plotone{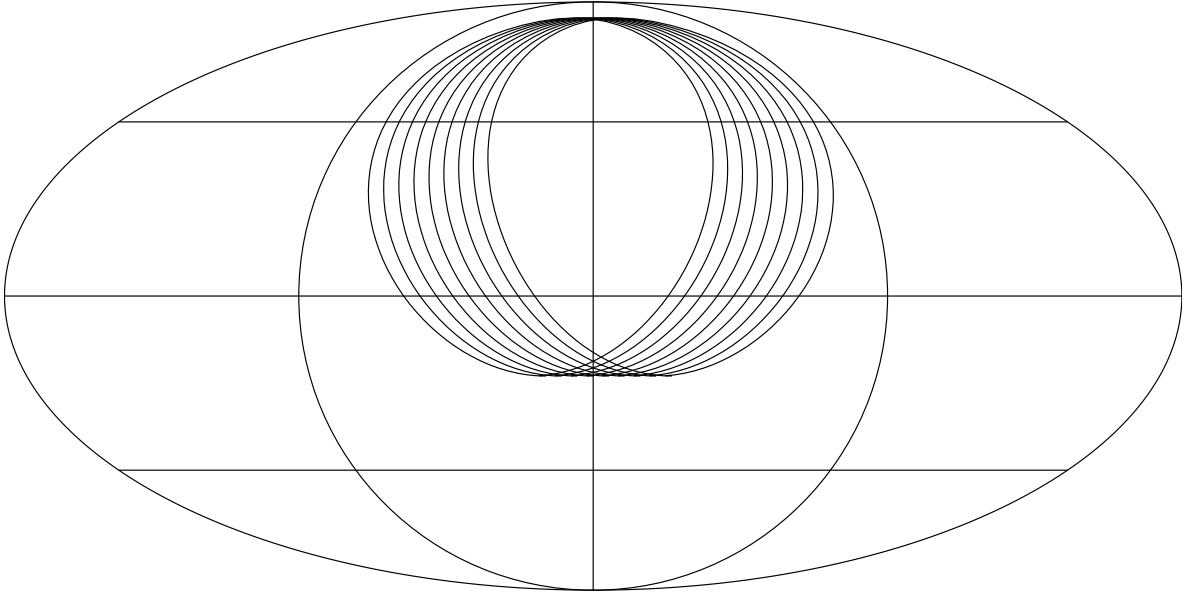}
\caption{A simple scan pattern generated by
rotation around the zenith and the diurnal rate.
\label{fig:firsscan.eps}}
\end{figure}

In terms of the matrices introduced earlier, only \mW\ needs to be 
changed for $1/f$ noise.  In the toy model, the scan circles are 
so short that only a very short pre-whitening filter can be used,
so I will use $W_k = -0.5, 1, \& -0.5$.  This gives
\be
\mW = \frac{1}{2}\left(
\begin{array}{rrrrrrrrrrrrrrrrrr}
   2 &  -1 &   0 &   0 &   0 &  -1 &   0 &   0 &   0 &   0 &   0 &   0 &   0 &   0 &   0 &   0 &   0 &   0 \\
  -1 &   2 &  -1 &   0 &   0 &   0 &   0 &   0 &   0 &   0 &   0 &   0 &   0 &   0 &   0 &   0 &   0 &   0 \\
   0 &  -1 &   2 &  -1 &   0 &   0 &   0 &   0 &   0 &   0 &   0 &   0 &   0 &   0 &   0 &   0 &   0 &   0 \\
   0 &   0 &  -1 &   2 &  -1 &   0 &   0 &   0 &   0 &   0 &   0 &   0 &   0 &   0 &   0 &   0 &   0 &   0 \\
   0 &   0 &   0 &  -1 &   2 &  -1 &   0 &   0 &   0 &   0 &   0 &   0 &   0 &   0 &   0 &   0 &   0 &   0 \\
  -1 &   0 &   0 &   0 &  -1 &   2 &   0 &   0 &   0 &   0 &   0 &   0 &   0 &   0 &   0 &   0 &   0 &   0 \\
   0 &   0 &   0 &   0 &   0 &   0 &   2 &  -1 &   0 &   0 &   0 &  -1 &   0 &   0 &   0 &   0 &   0 &   0 \\
   0 &   0 &   0 &   0 &   0 &   0 &  -1 &   2 &  -1 &   0 &   0 &   0 &   0 &   0 &   0 &   0 &   0 &   0 \\
   0 &   0 &   0 &   0 &   0 &   0 &   0 &  -1 &   2 &  -1 &   0 &   0 &   0 &   0 &   0 &   0 &   0 &   0 \\
   0 &   0 &   0 &   0 &   0 &   0 &   0 &   0 &  -1 &   2 &  -1 &   0 &   0 &   0 &   0 &   0 &   0 &   0 \\
   0 &   0 &   0 &   0 &   0 &   0 &   0 &   0 &   0 &  -1 &   2 &  -1 &   0 &   0 &   0 &   0 &   0 &   0 \\
   0 &   0 &   0 &   0 &   0 &   0 &  -1 &   0 &   0 &   0 &  -1 &   2 &   0 &   0 &   0 &   0 &   0 &   0 \\
   0 &   0 &   0 &   0 &   0 &   0 &   0 &   0 &   0 &   0 &   0 &   0 &   2 &  -1 &   0 &   0 &   0 &  -1 \\
   0 &   0 &   0 &   0 &   0 &   0 &   0 &   0 &   0 &   0 &   0 &   0 &  -1 &   2 &  -1 &   0 &   0 &   0 \\
   0 &   0 &   0 &   0 &   0 &   0 &   0 &   0 &   0 &   0 &   0 &   0 &   0 &  -1 &   2 &  -1 &   0 &   0 \\
   0 &   0 &   0 &   0 &   0 &   0 &   0 &   0 &   0 &   0 &   0 &   0 &   0 &   0 &  -1 &   2 &  -1 &   0 \\
   0 &   0 &   0 &   0 &   0 &   0 &   0 &   0 &   0 &   0 &   0 &   0 &   0 &   0 &   0 &  -1 &   2 &  -1 \\
   0 &   0 &   0 &   0 &   0 &   0 &   0 &   0 &   0 &   0 &   0 &   0 &  -1 &   0 &   0 &   0 &  -1 &   2 
\end{array}
\right)
\ee
The correlation matrix is
\be
\mA = \frac{1}{4}\left(
\begin{array}{rrrrrrrrrrrrrr}
  18 &  -4 &   1 &   1 &  -4 &  -4 &   1 &   1 &  -4 &  -4 &   1 &   1 &  -4 &   0 \\
  -4 &   6 &  -4 &   0 &   1 &   0 &   0 &   0 &   0 &   0 &   0 &   0 &   0 &   1 \\
   1 &  -4 &   6 &   1 &   0 &   0 &   0 &   0 &   0 &   0 &   0 &   0 &   0 &  -4 \\
   1 &   0 &   1 &   6 &  -4 &   0 &   0 &   0 &   0 &   0 &   0 &   0 &   0 &  -4 \\
  -4 &   1 &   0 &  -4 &   6 &   0 &   0 &   0 &   0 &   0 &   0 &   0 &   0 &   1 \\
  -4 &   0 &   0 &   0 &   0 &   6 &  -4 &   0 &   1 &   0 &   0 &   0 &   0 &   1 \\
   1 &   0 &   0 &   0 &   0 &  -4 &   6 &   1 &   0 &   0 &   0 &   0 &   0 &  -4 \\
   1 &   0 &   0 &   0 &   0 &   0 &   1 &   6 &  -4 &   0 &   0 &   0 &   0 &  -4 \\
  -4 &   0 &   0 &   0 &   0 &   1 &   0 &  -4 &   6 &   0 &   0 &   0 &   0 &   1 \\
  -4 &   0 &   0 &   0 &   0 &   0 &   0 &   0 &   0 &   6 &  -4 &   0 &   1 &   1 \\
   1 &   0 &   0 &   0 &   0 &   0 &   0 &   0 &   0 &  -4 &   6 &   1 &   0 &  -4 \\
   1 &   0 &   0 &   0 &   0 &   0 &   0 &   0 &   0 &   0 &   1 &   6 &  -4 &  -4 \\
  -4 &   0 &   0 &   0 &   0 &   0 &   0 &   0 &   0 &   1 &   0 &  -4 &   6 &   1 \\
   0 &   1 &  -4 &  -4 &   1 &   1 &  -4 &  -4 &   1 &   1 &  -4 &  -4 &   1 &  18
\end{array}
\right)
\ee
The covariance matrix is given by $\mA^{-1} = $
$$
{\scriptsize
\left(
\begin{array}{rrrrrrrrrrrrrr}
    0.54&   0.19&  -0.22&  -0.22&   0.19&   0.19&  -0.22&  -0.22&   0.19&   0.19&  -0.22&  -0.22&   0.19&  -0.38 \\
    0.19&   1.52&   0.93&  -0.85&  -0.70&  -0.01&  -0.20&  -0.20&  -0.01&  -0.01&  -0.20&  -0.20&  -0.01&  -0.22 \\
   -0.22&   0.93&   1.52&  -0.70&  -0.85&  -0.20&  -0.01&  -0.01&  -0.20&  -0.20&  -0.01&  -0.01&  -0.20&   0.19 \\
   -0.22&  -0.85&  -0.70&   1.52&   0.93&  -0.20&  -0.01&  -0.01&  -0.20&  -0.20&  -0.01&  -0.01&  -0.20&   0.19 \\
    0.19&  -0.70&  -0.85&   0.93&   1.52&  -0.01&  -0.20&  -0.20&  -0.01&  -0.01&  -0.20&  -0.20&  -0.01&  -0.22 \\
    0.19&  -0.01&  -0.20&  -0.20&  -0.01&   1.52&   0.93&  -0.85&  -0.70&  -0.01&  -0.20&  -0.20&  -0.01&  -0.22 \\
   -0.22&  -0.20&  -0.01&  -0.01&  -0.20&   0.93&   1.52&  -0.70&  -0.85&  -0.20&  -0.01&  -0.01&  -0.20&   0.19 \\
   -0.22&  -0.20&  -0.01&  -0.01&  -0.20&  -0.85&  -0.70&   1.52&   0.93&  -0.20&  -0.01&  -0.01&  -0.20&   0.19 \\
    0.19&  -0.01&  -0.20&  -0.20&  -0.01&  -0.70&  -0.85&   0.93&   1.52&  -0.01&  -0.20&  -0.20&  -0.01&  -0.22 \\
    0.19&  -0.01&  -0.20&  -0.20&  -0.01&  -0.01&  -0.20&  -0.20&  -0.01&   1.52&   0.93&  -0.85&  -0.70&  -0.22 \\
   -0.22&  -0.20&  -0.01&  -0.01&  -0.20&  -0.20&  -0.01&  -0.01&  -0.20&   0.93&   1.52&  -0.70&  -0.85&   0.19 \\
   -0.22&  -0.20&  -0.01&  -0.01&  -0.20&  -0.20&  -0.01&  -0.01&  -0.20&  -0.85&  -0.70&   1.52&   0.93&   0.19 \\
    0.19&  -0.01&  -0.20&  -0.20&  -0.01&  -0.01&  -0.20&  -0.20&  -0.01&  -0.70&  -0.85&   0.93&   1.52&  -0.22 \\
   -0.38&  -0.22&   0.19&   0.19&  -0.22&  -0.22&   0.19&   0.19&  -0.22&  -0.22&   0.19&   0.19&  -0.22&   0.54
\end{array}
\right)
}
$$

\section{Continuous Scanning}

The examples so far are based on scanning in discrete scan circles.
But continuous scanning is also possible.  With a continuous
scan \mW\ changes to
\be
\mW = \frac{1}{2}\left(
\begin{array}{rrrrrrrrrrrrrrrrrr}
   2 &  -1 &   0 &   0 &   0 &   0 &   0 &   0 &   0 &   0 &   0 &   0 &   0 &   0 &   0 &   0 &   0 &  -1\\
  -1 &   2 &  -1 &   0 &   0 &   0 &   0 &   0 &   0 &   0 &   0 &   0 &   0 &   0 &   0 &   0 &   0 &   0\\
   0 &  -1 &   2 &  -1 &   0 &   0 &   0 &   0 &   0 &   0 &   0 &   0 &   0 &   0 &   0 &   0 &   0 &   0\\
   0 &   0 &  -1 &   2 &  -1 &   0 &   0 &   0 &   0 &   0 &   0 &   0 &   0 &   0 &   0 &   0 &   0 &   0\\
   0 &   0 &   0 &  -1 &   2 &  -1 &   0 &   0 &   0 &   0 &   0 &   0 &   0 &   0 &   0 &   0 &   0 &   0\\
   0 &   0 &   0 &   0 &  -1 &   2 &  -1 &   0 &   0 &   0 &   0 &   0 &   0 &   0 &   0 &   0 &   0 &   0\\
   0 &   0 &   0 &   0 &   0 &  -1 &   2 &  -1 &   0 &   0 &   0 &   0 &   0 &   0 &   0 &   0 &   0 &   0\\
   0 &   0 &   0 &   0 &   0 &   0 &  -1 &   2 &  -1 &   0 &   0 &   0 &   0 &   0 &   0 &   0 &   0 &   0\\
   0 &   0 &   0 &   0 &   0 &   0 &   0 &  -1 &   2 &  -1 &   0 &   0 &   0 &   0 &   0 &   0 &   0 &   0\\
   0 &   0 &   0 &   0 &   0 &   0 &   0 &   0 &  -1 &   2 &  -1 &   0 &   0 &   0 &   0 &   0 &   0 &   0\\
   0 &   0 &   0 &   0 &   0 &   0 &   0 &   0 &   0 &  -1 &   2 &  -1 &   0 &   0 &   0 &   0 &   0 &   0\\
   0 &   0 &   0 &   0 &   0 &   0 &   0 &   0 &   0 &   0 &  -1 &   2 &  -1 &   0 &   0 &   0 &   0 &   0\\
   0 &   0 &   0 &   0 &   0 &   0 &   0 &   0 &   0 &   0 &   0 &  -1 &   2 &  -1 &   0 &   0 &   0 &   0\\
   0 &   0 &   0 &   0 &   0 &   0 &   0 &   0 &   0 &   0 &   0 &   0 &  -1 &   2 &  -1 &   0 &   0 &   0\\
   0 &   0 &   0 &   0 &   0 &   0 &   0 &   0 &   0 &   0 &   0 &   0 &   0 &  -1 &   2 &  -1 &   0 &   0\\
   0 &   0 &   0 &   0 &   0 &   0 &   0 &   0 &   0 &   0 &   0 &   0 &   0 &   0 &  -1 &   2 &  -1 &   0\\
   0 &   0 &   0 &   0 &   0 &   0 &   0 &   0 &   0 &   0 &   0 &   0 &   0 &   0 &   0 &  -1 &   2 &  -1\\
  -1 &   0 &   0 &   0 &   0 &   0 &   0 &   0 &   0 &   0 &   0 &   0 &   0 &   0 &   0 &   0 &  -1 &   2
\end{array}
\right)
\ee
and the correlation matrix changes to 
\be
\mA = \frac{1}{4}\left(
\begin{array}{rrrrrrrrrrrrrr}
  18 &  -4 &   1 &   1 &  -4 &  -4 &   1 &   1 &  -4 &  -4 &   1 &   1 &  -4 &   0\\
  -4 &   6 &  -4 &   0 &   0 &   0 &   0 &   0 &   0 &   0 &   0 &   0 &   1 &   1\\
   1 &  -4 &   6 &   1 &   0 &   0 &   0 &   0 &   0 &   0 &   0 &   0 &   0 &  -4\\
   1 &   0 &   1 &   6 &  -4 &   0 &   0 &   0 &   0 &   0 &   0 &   0 &   0 &  -4\\
  -4 &   0 &   0 &  -4 &   6 &   1 &   0 &   0 &   0 &   0 &   0 &   0 &   0 &   1\\
  -4 &   0 &   0 &   0 &   1 &   6 &  -4 &   0 &   0 &   0 &   0 &   0 &   0 &   1\\
   1 &   0 &   0 &   0 &   0 &  -4 &   6 &   1 &   0 &   0 &   0 &   0 &   0 &  -4\\
   1 &   0 &   0 &   0 &   0 &   0 &   1 &   6 &  -4 &   0 &   0 &   0 &   0 &  -4\\
  -4 &   0 &   0 &   0 &   0 &   0 &   0 &  -4 &   6 &   1 &   0 &   0 &   0 &   1\\
  -4 &   0 &   0 &   0 &   0 &   0 &   0 &   0 &   1 &   6 &  -4 &   0 &   0 &   1\\
   1 &   0 &   0 &   0 &   0 &   0 &   0 &   0 &   0 &  -4 &   6 &   1 &   0 &  -4\\
   1 &   0 &   0 &   0 &   0 &   0 &   0 &   0 &   0 &   0 &   1 &   6 &  -4 &  -4\\
  -4 &   1 &   0 &   0 &   0 &   0 &   0 &   0 &   0 &   0 &   0 &  -4 &   6 &   1\\
   0 &   1 &  -4 &  -4 &   1 &   1 &  -4 &  -4 &   1 &   1 &  -4 &  -4 &   1 &  18
\end{array}
\right)
\ee
The covariance matrix becomes $\mA^{-1} = $
$$
{\scriptsize
\left(
\begin{array}{rrrrrrrrrrrrrr}
  0.54&  0.19& -0.22& -0.22&  0.19&  0.19& -0.22& -0.22&  0.19&  0.19& -0.22& -0.22&  0.19& -0.38\\
  0.19&  1.38&  0.78& -0.51& -0.23&  0.06& -0.15& -0.23& -0.05&  0.06& -0.10& -0.51& -0.45& -0.22\\
 -0.22&  0.78&  1.38& -0.45& -0.51& -0.10&  0.06& -0.05& -0.23& -0.15&  0.06& -0.23& -0.51&  0.19\\
 -0.22& -0.51& -0.45&  1.38&  0.78& -0.51& -0.23&  0.06& -0.15& -0.23& -0.05&  0.06& -0.10&  0.19\\
  0.19& -0.23& -0.51&  0.78&  1.38& -0.45& -0.51& -0.10&  0.06& -0.05& -0.23& -0.15&  0.06& -0.22\\
  0.19&  0.06& -0.10& -0.51& -0.45&  1.38&  0.78& -0.51& -0.23&  0.06& -0.15& -0.23& -0.05& -0.22\\
 -0.22& -0.15&  0.06& -0.23& -0.51&  0.78&  1.38& -0.45& -0.51& -0.10&  0.06& -0.05& -0.23&  0.19\\
 -0.22& -0.23& -0.05&  0.06& -0.10& -0.51& -0.45&  1.38&  0.78& -0.51& -0.23&  0.06& -0.15&  0.19\\
  0.19& -0.05& -0.23& -0.15&  0.06& -0.23& -0.51&  0.78&  1.38& -0.45& -0.51& -0.10&  0.06& -0.22\\
  0.19&  0.06& -0.15& -0.23& -0.05&  0.06& -0.10& -0.51& -0.45&  1.38&  0.78& -0.51& -0.23& -0.22\\
 -0.22& -0.10&  0.06& -0.05& -0.23& -0.15&  0.06& -0.23& -0.51&  0.78&  1.38& -0.45& -0.51&  0.19\\
 -0.22& -0.51& -0.23&  0.06& -0.15& -0.23& -0.05&  0.06& -0.10& -0.51& -0.45&  1.38&  0.78&  0.19\\
  0.19& -0.45& -0.51& -0.10&  0.06& -0.05& -0.23& -0.15&  0.06& -0.23& -0.51&  0.78&  1.38& -0.22\\
 -0.38& -0.22&  0.19&  0.19& -0.22& -0.22&  0.19&  0.19& -0.22& -0.22&  0.19&  0.19& -0.22&  0.54
\end{array}
\right)
}
$$
which is slightly better than the scan circle case because the continuous scan 
introduces some new comparisons.

For large datasets and large maps, the actual construction of \mA\ and
$\mA^{-1}$ is impractical.  However, the evaluation of \mA\mX\ for any
vector \mX\ can be done in a reasonable amount of time by following
these steps:
\bn
\item Apply \mP: Fly the mission through the map, which takes
${\cal O}(n_d)$ operations.
\item Apply $\mW^T\mW$: This is a convolution.  For the scan circle
case, the convolution is done in $N$ sample blocks, and the total work is 
${\cal O}(n_d \ln N)$ operations when using the FFT.  
For the continuously scanned case, the filter to be applied to the data is 
${\cal O}(L_W)$ points long and the total work is
${\cal O}(n_d \ln L_W)$ when using FFTs.
\item Apply $\mP^T$: Sum into pixels, which takes
${\cal O}(n_d)$ operations.
\en
Thus the evaluation of \mA\mX\ takes ${\cal O}(n_d \ln L_W)$ operations
and standard methods for the iterative solution of sparse systems that only
need matrix-vector products can be used.
This process should converge well for a complex scan pattern that
crosses each pixel several times in different directions, such as
the cycloidal scan produced by the scan plate in Lubin's balloon
experiment, shown in Figure \ref{fig:ucsbscan.eps}.
A simple scan pattern such as rotating at 
constant elevation while the Earth turns,
shown in Figure \ref{fig:firsscan.eps},
will lead to much slower
convergence of the iterations.  
Since pixels are always scanned in 
the same direction (unless the flight is long enough to see the sky
both rising and setting), the simple scan pattern
will give a final map with stripes.

\section{Bigger Examples}

\begin{figure}[tbp]
\plotone{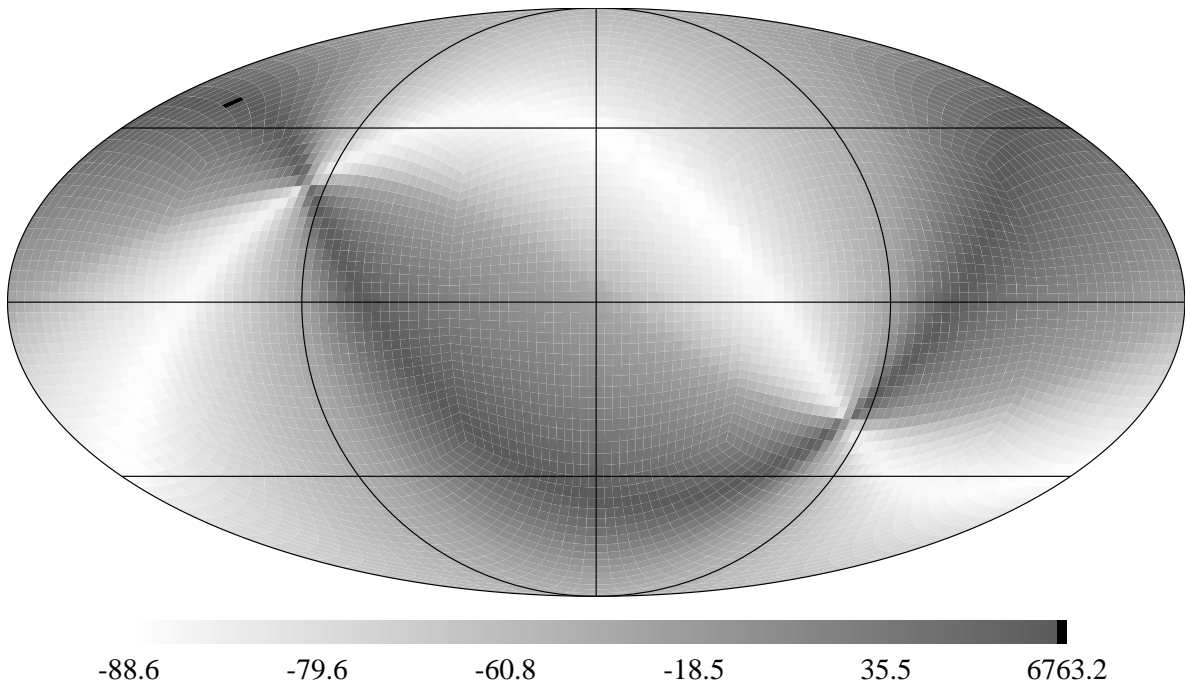}
\caption{The row through the Lockman Hole of the correlation matrix \mA\ and 
the covariance matrix $\mA^{-1}$ for the white noise case.
\label{fig:spinner0.eps}}
\end{figure}

\begin{figure}[tbp]
\plotone{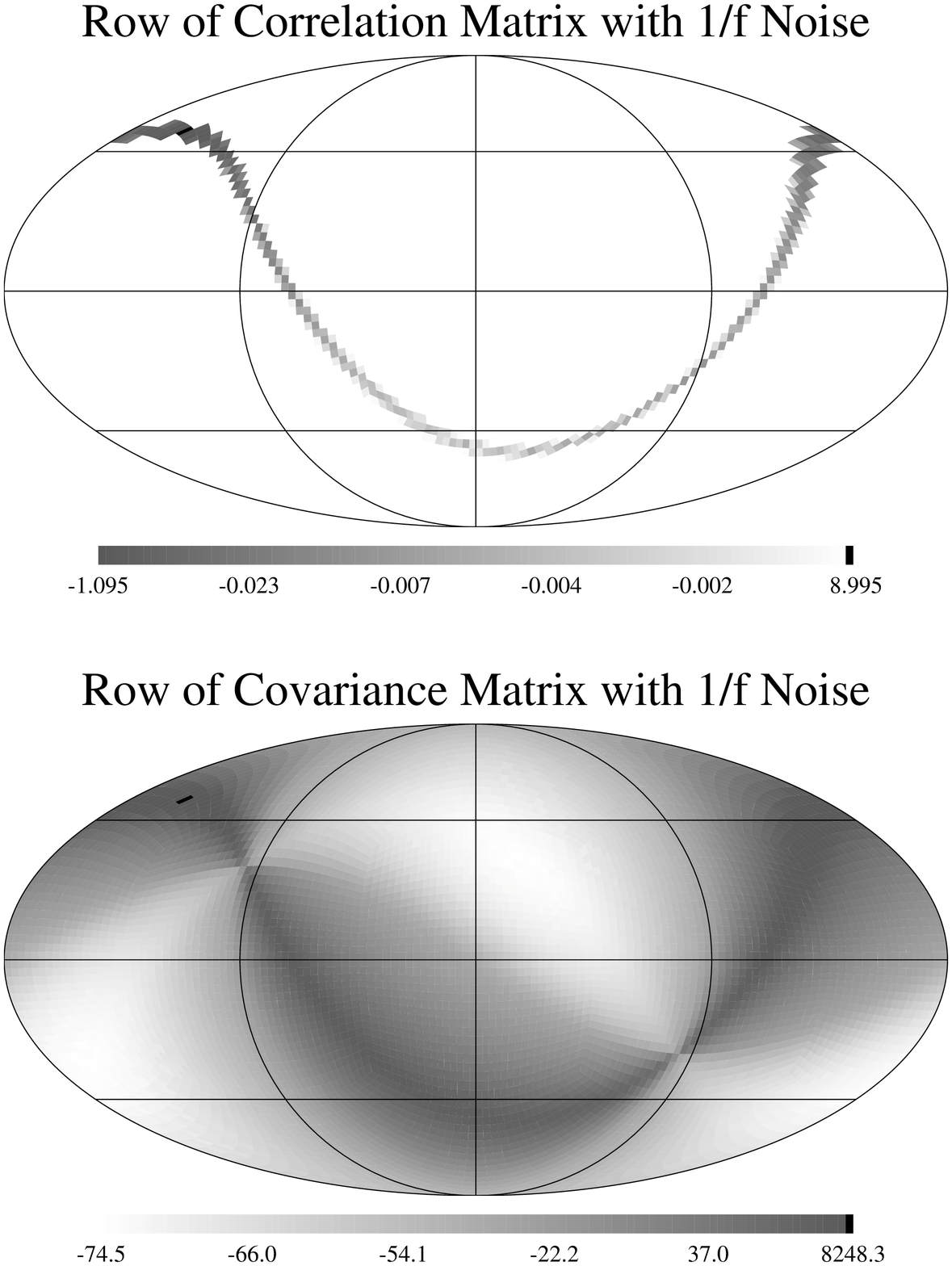}
\caption{The row through the Lockman Hole of the correlation matrix \mA\ and 
the covariance matrix $\mA^{-1}$ for the $1/f$ noise case.
The $1/f$ knee is at $f_K = 10 f_{spin}$ so the effective chop angle is
$13.4^\circ$.
\label{fig:spinner1.eps}}
\end{figure}

\begin{figure}[tbp]
\plotone{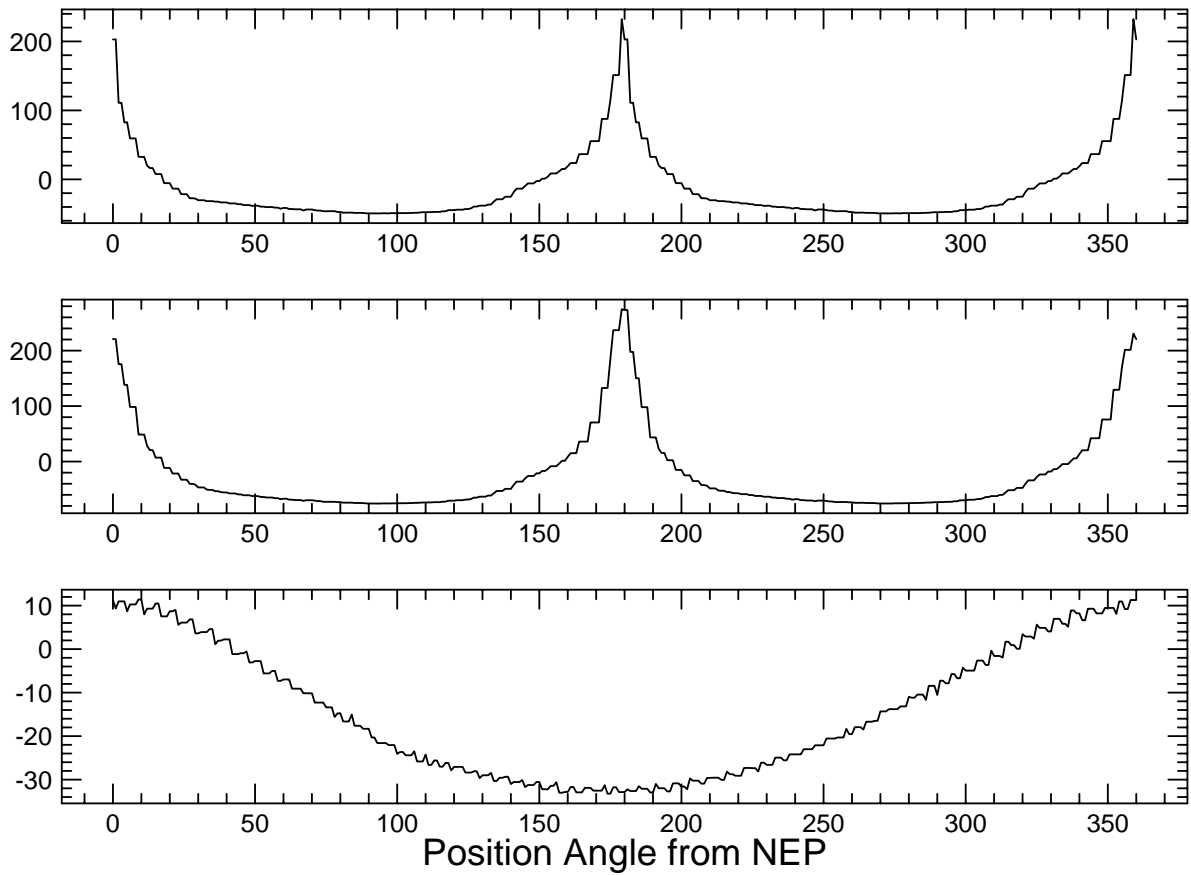}
\caption{The covariance of pixels $90^\circ$ from the Lockman Hole
as a function of position angle with respect to the NEP
for 
the white noise case (top) and  
the $1/f$ noise case (middle).
Note
the increased covariance on the scan circle through the Lockman hole
at $0^\circ$ and $180^\circ$.
Bottom: the COBE scan case, with $1/f$ noise.  Note the absence of
stripes.
\label{fig:cvslong.eps}}
\end{figure}

These equations have been implemented for a bigger example using DMR pixels
in galactic coordinates, and 256 scans through the ecliptic poles evenly spaced
in ecliptic longitude, each scan having 256 points.  Thus there are a total
of $n_d = 256^2 = 65536$ observations for $n_p = 6144$ pixels.  Since the
whole matrices \mA\ or $\mA^{-1}$ are too big to print out, I have made maps
from a single row of these matrices.  The pixel on the diagonal in this row
is the Lockman Hole at $l = 150.5^\circ$ and $b = 53^\circ$.  It was observed
10 times, which is close to the average exposure for the map.
Figure \ref{fig:spinner0.eps} shows the white noise case.  There is a small
amount of excess noise and striping.
The covariance matrices have been multiplied by the total number of
data points, so a perfect experiment would have $n_p$ on the diagonal
and $-1$ in the off-diagonal pixels.
Figure \ref{fig:spinner1.eps} shows the $1/f$ noise case.  There is a larger
amount of excess noise and striping, but it is concentrated in the low
order multipoles.
The angle scanned during one radian at $f_K$ for this case is $5.7^\circ$,
so excess noise in low multipoles is expected.  Half of the weight in the
baseline filter in Figure \ref{fig:w_0p1.eps} is contained within
$\pm 13.4^\circ$ of the central pixel, so $13.4^\circ$ is the
effective chop angle.
Figure \ref{fig:cvslong.eps} shows the covariance on a circle $90^\circ$
away from the central pixel, which peaks sharply when crossing the
scan circle -- an indication of striping.

\begin{figure}[tbp]
\plotone{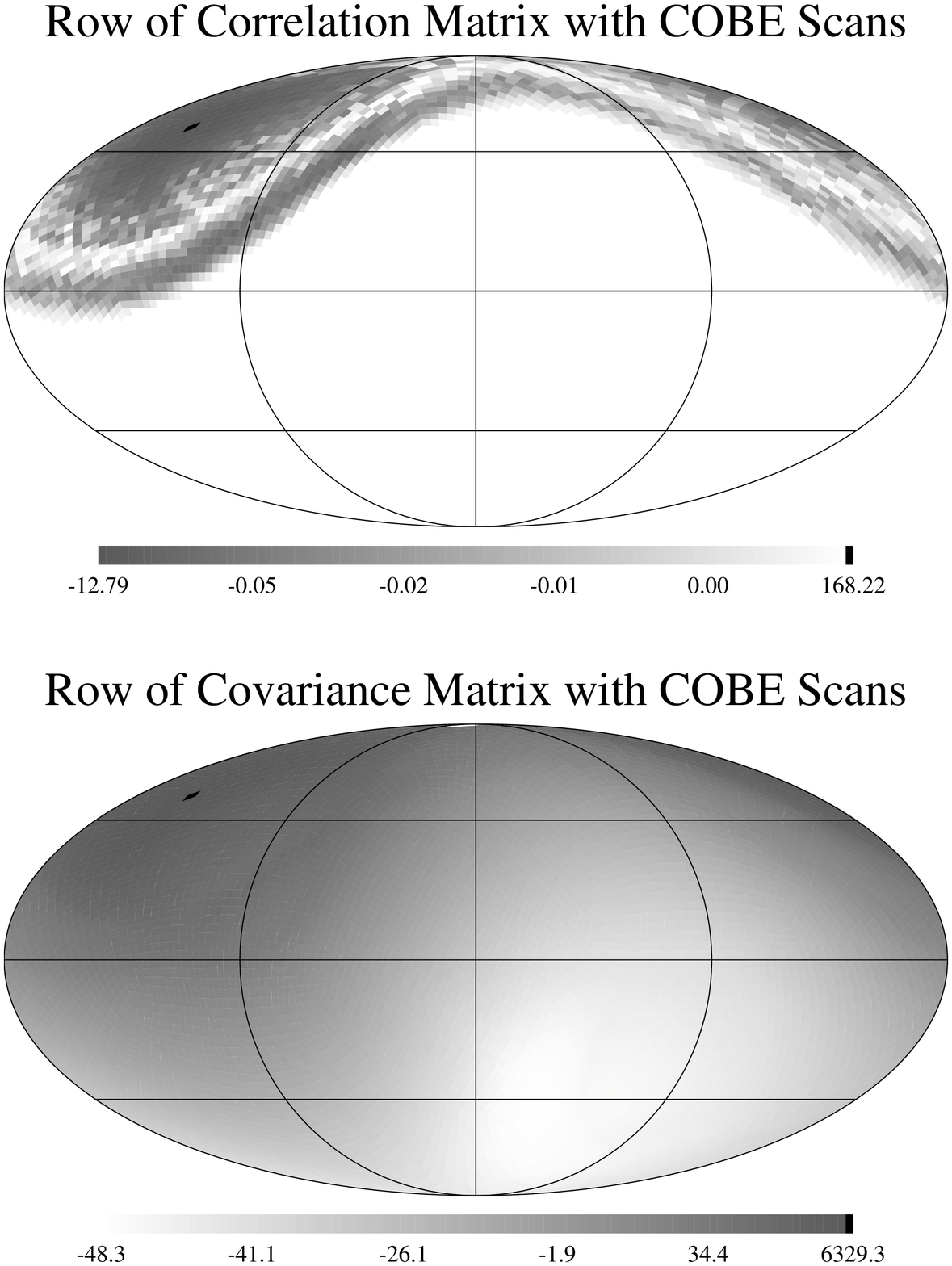}
\caption{The row through the Lockman Hole of the correlation matrix \mA\ and 
the covariance matrix $\mA^{-1}$ for the COBE scan case with
$1/f$ noise.
The $1/f$ knee is at $f_K = 5.5 f_{spin}$ so the effective chop angle
is $11.8^\circ$.
\label{fig:cobe1.eps}}
\end{figure}

An even bigger example (Figure \ref{fig:cobe1.eps})
shows the result of a one-horned experiment using the
{\sl COBE} DMR scan path with $1/f$ noise.  
One million data points were processed, and the
orbit precession (yearly) rate was increased by a factor of 60
so these $10^6$ points covered a full annual cycle.
The length of the pre-whitening filter was $L_W = 100$.  
Slow convolutions instead
of FFT's were used to evaluate a row of \mA\ and to iterate to find a row 
of $\mA^{-1}$.  The time-ordered method took 30 seconds on a workstation
to evaluate a product of the form $\mA\mT$.  
The residuals in $\mA\mT = \mB$ were reduced by a factor $10^{-5}$ after
20 iterations of the conjugate gradient method.
These results are plotted at DMR
resolution but the timing is independent of the number of pixels, $n_p$, and
directly proportional to the number of data points, $n_d$.  

The angle scanned in one radian at $f_K$ was $5.0^\circ$, so the only advantage
this case has over Figure \ref{fig:spinner1.eps} is that scans were made
through each pixel in every direction.  Thus it is not necessary to walk the
baseline determination up to the NEP and then down along some other longitude.
Thus the final covariance depends primarily on the distance between two points,
giving a much more symmetric pattern in Figure \ref{fig:cobe1.eps}.
The Lockman Hole was observed 189 times, and the diagonal element of the
covariance matrix is 0.00633, so there is 20\% excess noise at the pixel level.
The highest off-diagonal element in the Lockman Hole row is 8 times smaller
than the diagonal element.

In general a symmetric covariance is much easier to analyze, since it leads
to a noise level that depends only on $\ell$.  But in the absence of
systematic errors,
Figures \ref{fig:spinner0.eps} and even \ref{fig:spinner1.eps} can give useful
CMB data -- especially if the overall noise level is low enough
(Janssen \etal\markcite{JSWSL96} 1996).

\section{Systematic Errors}

\begin{figure}[tbp]
\plotone{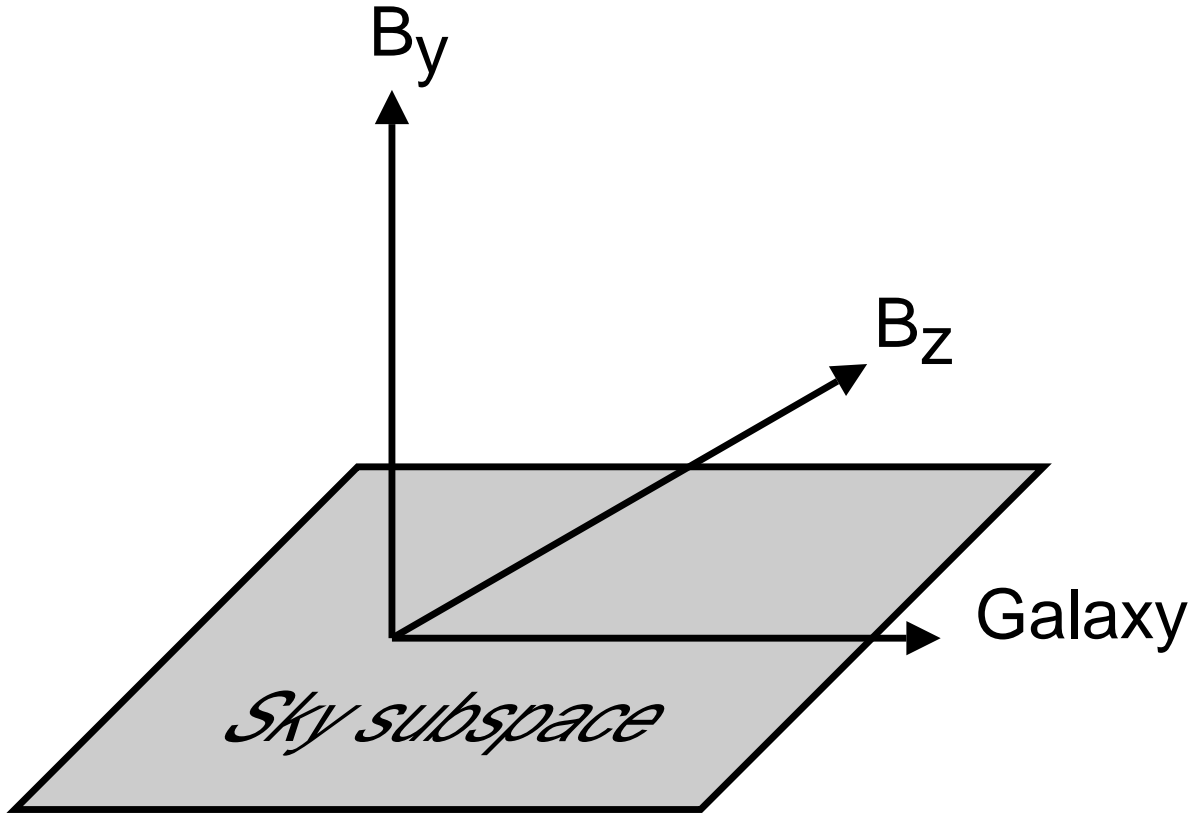}
\caption{A diagram showing the $n_d$-dimensional ``data space'' and the
$n_p$-dimensional subspace of sky patterns.  A given systematic error
is a direction in data space that can be also orthogonal to the sky subspace,
like the $B_y$ magnetic effect, almost but not quite parallel to the
sky subspace like the $B_z$ magnetic effect, or completely contained within the
sky subspace like a galactic foreground (for 1 frequency maps).
\label{fig:sys_err.eps}}
\end{figure}

\begin{figure}[tbp]
\plotone{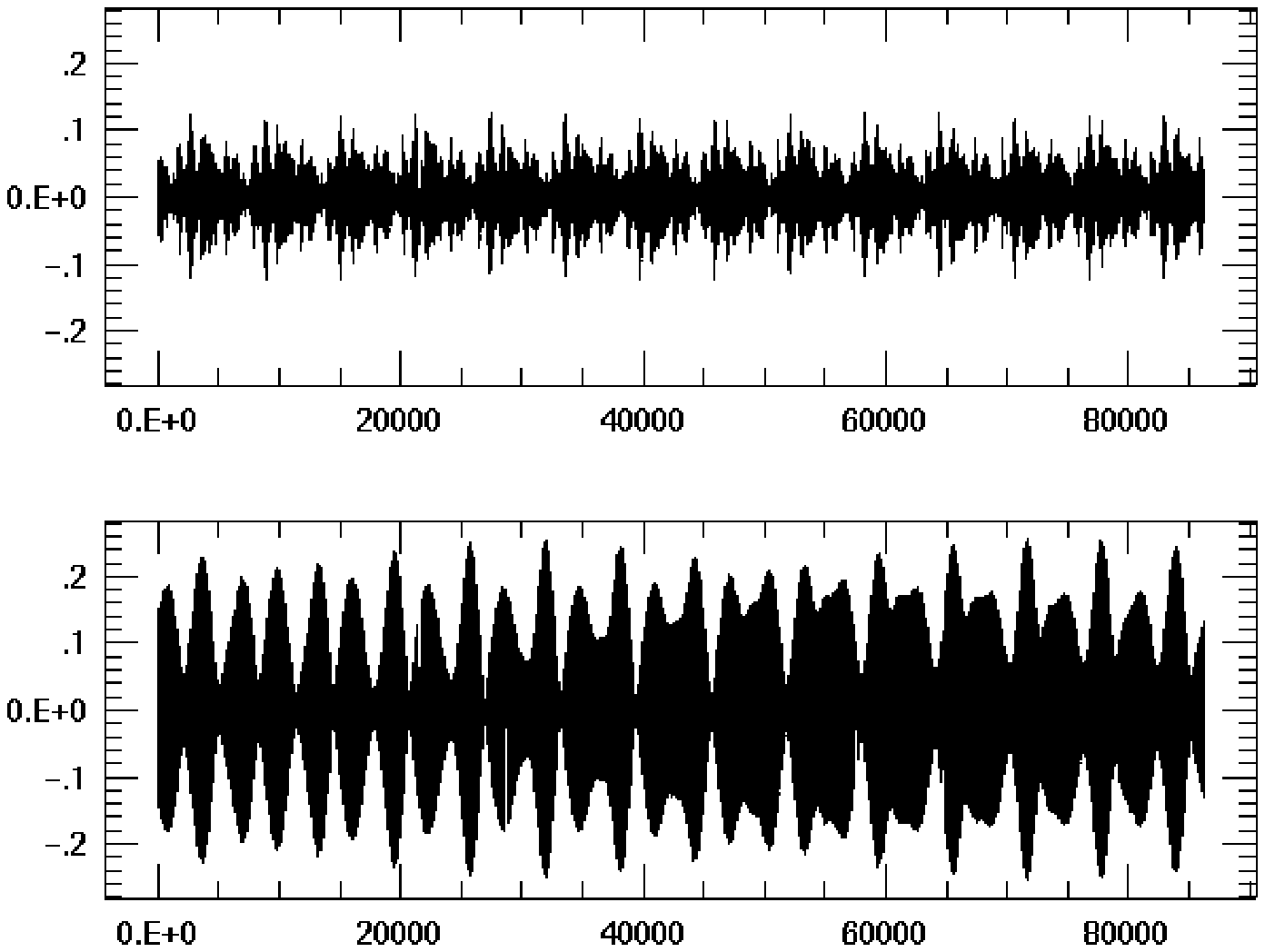}
\caption{Bottom: Time-ordered signal in mK \vs\ seconds
from a 1 mK/g $y$-axis magnetic 
sensitivity in the DMR.  Top: Time-ordered signal resulting from a 
playback of the map created from the bottom input.
\label{fig:bysyserr.eps}}
\end{figure}

\begin{figure}[tbp]
\plotone{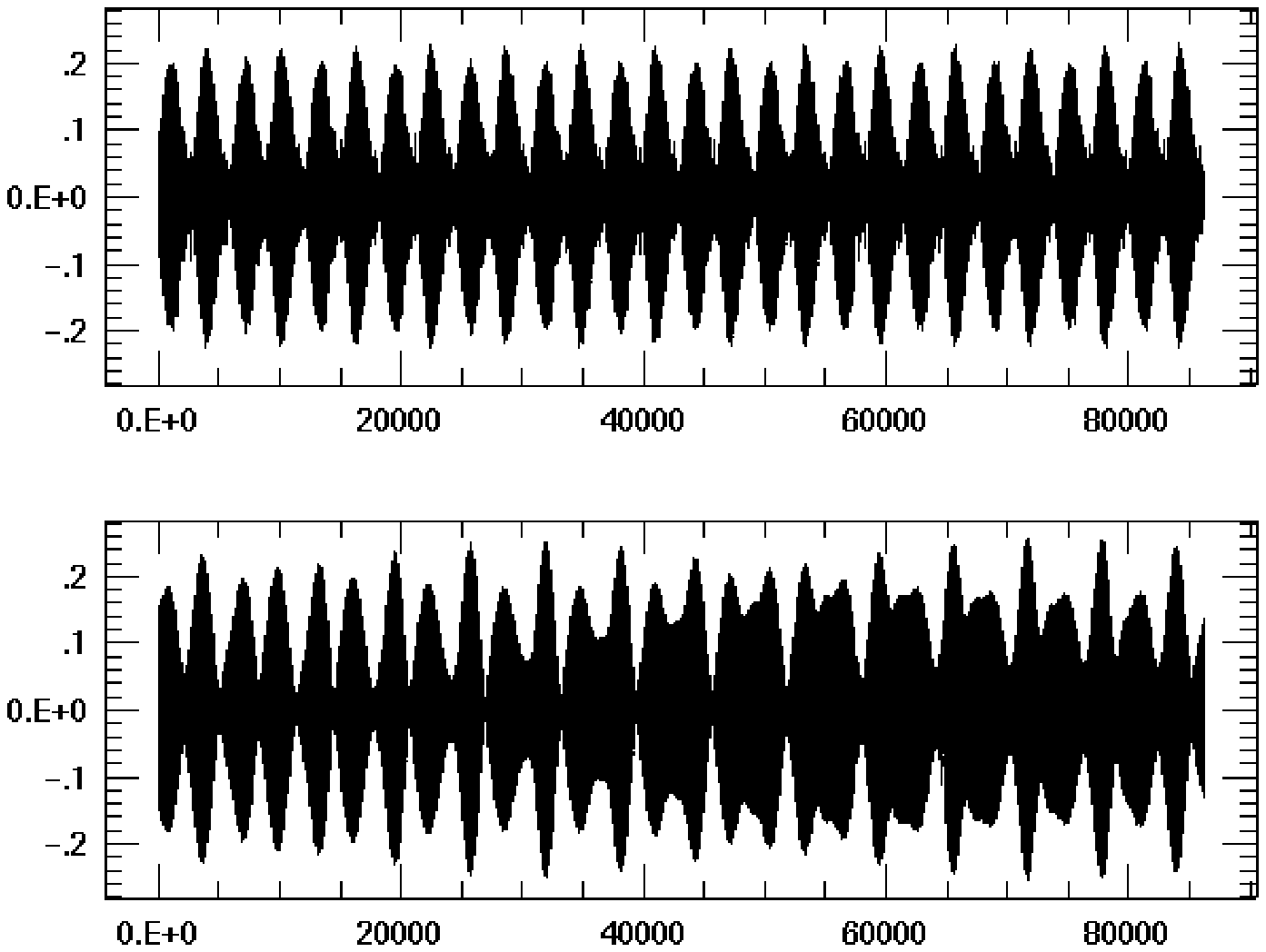}
\caption{Bottom: Time-ordered signal from a 1 mK/g $z$-axis magnetic 
sensitivity in the DMR.  Top: Time-ordered signal resulting from a 
playback of the map created from the bottom input.
\label{fig:bzsyserr.eps}}
\end{figure}

A systematic error term is a non-random signal in the time-ordered data that
is not caused by the sky.  Since true patterns on the sky form an
$n_p$-dimension subspace in the $n_d$-dimensional data space, a systematic
error term can be
\bn
\item Orthogonal to the sky subspace - these can be ignored.
\item Parallel to the sky subspace - these cannot be fixed.
\item At an intermediate angle to the sky subpsace - these must be
measured and corrected.
\en
Figure \ref{fig:sys_err.eps} shows these possibilities.

An example of a systematic error that is almost orthogonal to the
sky subspace is the $B_y$ magnetic term in the {\sl COBE} DMR.
This is a susceptibility to a magnetic field along the axis perpendicular to
the plane containing the plus and minus horns on a DMR.
The bottom plot in
Figure \ref{fig:bysyserr.eps} shows the time-ordered data produced by this term,
which can then be made into a map.  That map is then converted back into
time-ordered data, giving the top plot.  Since the top and bottom plots are
very different, it is easy to measure the $B_y$ sensitivity and correct for it.

A $B_z$ term is due to a sensitivity to magnetic fields in the plane defined by
the horns but perpendicular to the spin axis.  This produces a map with a large
dipole aligned with the celestial pole.  But the $B_z$ term is not completely
parallel to the sky subspace because the Earth's magnetic dipole is offset and
tilted so there is a diurnal modulation of the input shown on the
bottom of Figure \ref{fig:bzsyserr.eps} that is not seen in the playback (top).
Thus it is possible to measure a $B_z$ term and remove its effect from the map.

An example of a systematic error that is parallel to the sky subspace for
a simple scan pattern perpendicular to the Sun-line through the ecliptic poles
is zodiacal light emission.  With the more complex scan pattern used by the
DMR this effect can be measured and removed from the map.

\section{Desiderata}

Obviously one should try to make all systematic errors orthogonal to the sky
subspace.  This is made easier if the data space is much larger than the sky
subspace.  Therefore the number of data points transmitted to the ground
should be as large as possible, and on-board co-addition into scan circles
should be avoided.

To avoid stripes, scans should be made through every pixel in all possible
directions.  This requires a large number of scans and a large number of data
points, which are also desired to reduce systematic error problems.

\section{Conclusion}

I have shown how to analyze large datasets from one-horned CMB experiments
without simplified scan patterns that are vulnerable to systematic errors.
The CPU time required is ${\cal O}(n_d \ln L_W)$ is differs from the
Wright \etal\markcite{WHB96} (1996) 
differential method timing only in a logarithmic factor.

\end{document}